\documentstyle[12pt,epsfig]{article}
\newcommand{\be}{\begin{equation}}
\newcommand{\ee}{\end{equation}}

\begin{document}

\begin{titlepage}
\begin{center}
{\hbox to\hsize{
\hfill MIT-CTP-2892}} 

{\hbox to\hsize{ 
\hfill PUPT-1883}}

{\hbox to\hsize{ 
\hfill NSF-ITP-99-092}}

\bigskip
\vspace{3\baselineskip}

{\Large \bf  
 
The Shape of Gravity\\}

\bigskip

\bigskip

{\bf Joseph Lykken}\\
\smallskip
{\small \it Institute for Theoretical Physics,\\ 
University of California, Santa Barbara, CA 93106.\\

and

Theoretical Physics Dept.\\
Fermi National Accelerator Laboratory, Batavia, IL 60510.}

\medskip

{\bf  Lisa Randall}\\
\smallskip

{ \small \it  
Joseph Henry Laboratories,
Princeton University,
Princeton, NJ 08543, USA\\

and

Center for Theoretical Physics,\\
Massachusetts Institute of Technology, Cambridge, MA 02139, USA }

\bigskip

{\tt lykken@fnal.gov}\\
{\tt randall@feynman.princeton.edu}

\bigskip

\vspace*{.3cm}

{\bf Abstract}\\
\end{center}
\noindent
In a nontrivial background geometry with extra dimensions,
gravitational effects will depend on the shape of the
Kaluza-Klein excitations of the graviton. 
We investigate a consistent scenario of this type
with two positive tension three-branes separated in
a five-dimensional Anti-de Sitter geometry. The graviton is
localized on the ``Planck'' brane, while a gapless continuum
of additional gravity eigenmodes probe the {\it infinitely} large
fifth dimension. Despite the
background five-dimensional geometry,
an observer confined to either brane sees gravity
as essentially four-dimensional up to a
position-dependent strong coupling scale, no matter where
the brane is located. We apply this scenario
to generate the TeV scale as a hierarchically
suppressed mass scale. Arbitrarily light
gravitational modes appear in this scenario, but 
with suppressed couplings.  Real emission of these modes is observable
at future colliders; the effects are similar to those produced
by {\it six} large toroidal dimensions.

\bigskip

\bigskip

\end{titlepage}

Extra dimensions provide an alternative route
to addressing the hierarchy problem. This is because
the Planck scale, describing the strength of the
graviton coupling at low energies, is a derived scale.  In a simple
factorizable geometry, the Planck scale of a four-dimensional
world is related to that of a higher dimensional
world simply by a volume factor. The
large Planck scale indicates weak
graviton coupling which is in turn
a consequence of the large volume
over which the graviton can propagate \cite{add1}. In
this scenario, a large hierarchy
only arises in the presence
of a large volume for the compactified
dimensions, which is very difficult
to justify.  A more compelling alternative
has been suggested in Ref. \cite{rs2}. 
The idea of this paper was that the weak
graviton coupling arises because of an
interesting shape of the graviton wave
function in the extra dimensions. The
graviton is localized away from the 3+1-dimensional
world on which the Standard Model resides.
The large value of the Planck scale
arises because of the small amplitude
for the graviton to coincide with our ``brane''.

In Ref. \cite{rs3}, it was shown that the geometry
of a single brane with cosmological energy densities
tuned to guarantee Poincare invariance takes the form:
\begin{equation}
ds^2=e^{-2 k |y|} \eta_{\mu \nu} dx^\mu dx^\nu+dy^2 \quad ,
\end{equation}
where $\mu,\nu$ parameterize the four-dimensional coordinates
of our world, and $y$ is the coordinate of a fifth dimension.
The remarkable aspect of the above geometry is that
it gives rise to a localized graviton field. Mechanisms
for confining matter and gauge fields to a smaller
dimensional subspace were already known. The
new feature here is that the background geometry
gives rise to a single gravitational bound state. 
This mode plays the role of the graviton
of a four-dimensional world, and is responsible
for reproducing four-dimensional gravity. In Ref. \cite{rs3},
the Kaluza-Klein (KK) spectrum reflecting the large
extra dimension was derived and it was argued
that the additional {\it continuum gapless} spectrum
of states gives rise to very suppressed corrections
to conventional four-dimensional gravity, suppressed
by $({\rm energy}/M_{Pl})^2$.

However, from the perspective
of generating the mass hierarchy between
the Planck and weak scales,
the important aspect of this geometry
is the correspondence
between location in the fifth dimension and the overall
mass scale. This can be understood by the
fact that the warp factor is a conformal factor
so far as a four-dimensional world located
at a fixed $y$ location is concerned.  Mass factors
are rescaled by this factor, so that a natural
scale for mass parameters might be $M_{Pl}=10^{19}$ GeV
on a brane at the origin, but is $M_{Pl}$exp$-{k |y|}$ for
physics confined to a location $y$. This {\it exponential}
could be the source of the hierarchy between
the electroweak scale of order TeV and the Planck scale
which is approximately $10^{15}$ times bigger.
Notice that the generation of this hierarchy only requires
an exponential of order 30.

In Ref. \cite{rs2}, this observation was exploited
by introducing an orbifold geometry, and located
a positive energy brane at one point and a negative
energy brane at the second orbifold point. If the
standard model is located on this second, negative
energy brane, the amplitude of the graviton
is exponentially suppressed and  a hierarchy is generated.

The potential disadvantage of this setup
is the necessity for the negative energy object
and the orbifold geometry. Although not
ruled out, it is desirable to have an
alternative setup involving only positive
energy objects. The advantages
of such a setup are as follows. First,
there are positive energy objects,
namely D-branes and NS-branes,
that are well understood and on
which gauge fields and matter
fields can be localized so
that the Standard Model fields
can be placed there. Second, some potentially problematic
aspects of the cosmology of this system
were presented in \cite{csaki,kaloper}, though
it is not yet clear how general
the conclusions will prove.
Finally, there is the aesthetic advantage
of allowing for an infinite dimensional
space in which mass scales are associated
with definite locations in the space,
a point further emphasized by \cite{verlinde}.
If one permits all possible mass scales (all
possible distances in the fifth dimension), 
one presumably has a better chance of
addressing difficult cosmological issues
such as the cosmological constant
problem and black-hole physics. One
also has a better chance for exploiting
holographic ideas by exploiting the correspondence
between location in the fifth dimension and
mass scale.

In this paper, we demonstrate
that one can address
the hierarchy problem
with only positive energy objects
by combining the two observations
of Ref. \cite{rs2,rs3}, namely
1) it is consistent to live
with an infinite fifth dimension,
and 2) one can generate a hierarchy
by living far from the brane
on which gravity is localized.
This was implicit also in  Ref. \cite{verlinde},
where the connection between distance in the fifth dimension
and overall mass scale was made explicit in an AdS geometry
derived from D-3 branes (so that the Maldacena conjecture \cite{maldacena}
could be exploited), and so that the TeV scale
corresponded to a fixed coordinate $y_0$.

The crucial question is whether an observer on this ``TeV brane''
sees a consistent theory of gravity. In Ref. \cite{rs3},
it was only shown that 
one sees a theory of gravity
that is very close to a four-dimensional
gravitational theory if one
lives on the brane on which the
graviton is localized. In this paper,
we argue that even for an observer
quite far from that brane, one
obtains an acceptable gravitational
theory, essentially indistinguishable
from a four-dimensional world!

The picture that emerges
is remarkably beautiful. The graviton
is localized on a brane that we
call the Planck brane from now on.
We live on a brane separated from the Planck brane by roughly
30 Planck lengths in the fifth dimension.
On this brane, mass scales
are exponentially suppressed, yielding
a natural generation of the weak scale.
Furthermore, the maximal scale we
can probe on our brane is this same TeV
scale, since all string modes
become strongly interacting at this scale.
The location of the brane, which we denote
by $y_0$, was determined to give
the correct ratio of the weak scale to the Planck scale.
We call the brane at this location the TeV brane.

The potentially dangerous aspect 
of this setup is the multiplicity of the 
arbitrarily light Kaluza-Klein modes.
In Ref. \cite{rs2}, it was argued that
the KK modes were 
extremely strongly
coupled (with TeV coupling suppression rather than $M_{Pl}$).
In Ref. \cite{rs3}, it was shown that 
one signal of the infinite extra dimension
is a gapless continuum of Kaluza-Klein modes.
Clearly, if these modes were all so strongly
coupled, the theory would be disastrous,
since gravitational and particle
physics tests would be badly violated.

What we show in this paper is
that the situation is far more clever.
Production of modes less than the TeV
scale are suppressed.  Futhermore, modes
less than $10^{-4}$ eV
(which happens to correspond to the length scales on which
gravity has been directly probed,)
still couple with Planck scale suppression.
Thus the theory interpolates between 
a four-dimensional and five-dimensional
world (reminiscent of a holographic interpretation).
The observer on the brane at the TeV scale
sees the modes below a TeV in energy as weakly coupled. Modes
higher in mass than a TeV are much
more strongly coupled, and would
in principle reproduce the expected
five-dimensional result.
However, they are impossible to access!
Generalizing to an arbitrary location,
one never recognizes the higher dimensional
geometry. Independent of location,
the world appears lower dimensional
at low energies.

We now elaborate on this observation.
The results follow readily from
papers \cite{rs2,rs3}.
Our setup is a 
``Planck brane'' (or set of branes) on which the
graviton zero mode is confined, exponentially
falling off in the direction $y$. The new
feature
is a single brane (or multiple branes) located
a distance $y_0$ from this brane,
where $e^{-k y_o}={\rm TeV}/M_{Pl}$, where $k$
is related to the cosmological constant
on the brane and determines the exponential
falloff of the graviton, as in Ref. \cite{rs2,rs3}.
The new brane can be regarded as a probe of the geometry
determined by the Planck brane, either by assuming that
the Planck brane has much larger tension, or consists of a
large set of branes. It is readily seen that inclusion
of a small brane tension does not substantially
affect the result. We also remark that we do not address
the question of determining the location $y_0$ here,
though mechanisms that stabilize the orbifold
geometry (such as in Ref. \cite{wise}) should
also apply.

It is clear that the zero mode generates
consistent gravity. If we take the coordinate
$y=0$ to be the location of the Planck brane,
one can readily derive: 
\begin{equation}
M_{Pl}^2=2\int_0^\infty dy\; e^{-2 k y} M^3={M^3 \over k}\quad ,
\end{equation}
so that with $M$ and $k$ taken of order $M_{Pl}=10^{19}$
GeV, the zero mode is coupled correctly to generate
four-dimensional gravity.

It is therefore the contribution of
the additional KK modes that is our focus.
Everything follows from the detailed
form of these modes, derived in \cite{rs3}.
The graviton zero mode (properly normalized) is
\begin{equation}
\label{zerom}
\hat{\Psi}_0(z) = {1\over k(|z|+1/k)^{3/2}} \quad ,
\end{equation}
where the coordinate $z$ is related to $y$ by the expression
\be
z = {{\rm sgn}(y)\over k}\left( e^{k|y|} - 1 \right) \quad .
\ee
Note that at the TeV brane $z=z_0 \sim 1$ TeV$^{-1}$.

The continuum KK modes are given by:
\begin{equation}
\label{eigen}
\hat{\psi}_m \sim N_m
(|z|+1/k)^{1/2}\left  [Y_2(m  (|z|+1/k))  +  {4 k^2 \over \pi m^2}
J_2(m  (|z|+1/k)) \right ] \quad ,
\end{equation}
where $m$ is the mass of the mode, $Y_2$ and $J_2$ are Bessel functions,
and $N_m$ is a normalization factor.

For large $mz$, these modes asymptote to continuum plane wave behavior.
This can be seen from the asymptotic form for the Bessel functions:
\begin{equation}
\label{largez}
\sqrt{z} J_2(m z) \sim \sqrt{{2 \over \pi m }} \cos(mz -{5 \over 4} \pi)\; , 
~~~~ 
\sqrt{z} Y_2(m z) \sim \sqrt{{2 \over \pi m }} \sin(mz -{5 \over 4} \pi)\; .
\end{equation}
The normalization constants $N_m$ are determined by
this plane wave behavior \cite{rs3}:
\be
N_m \sim {\pi m^{5/2}\over 4 k^2} \quad .
\ee
We are adopting here a delta-function normalization such that
physical quantities will always involve an integration over
$m$ for which the proper measure is just $dm$. None of our
calculations will involve any dependence on the $y\rightarrow\infty$
regulator scheme (i.e. the ``regulator brane'' of \cite{rs3} or
the alternative proposed in \cite{verlinde}).

It is edifying to consider what these modes
tell us in a couple of limiting situations.
First, let us remind the reader
of what happens if you live on the Planck
brane ($z=0$). The exact effect
depends on the particular gravitational
process under consideration; let us
first consider the corrections to Newton's law from
the KK modes.
One finds a potential  between two masses $m_1$ and $m_2$:
\begin{equation}
\label{newtonsmall}
V=G_N{m_1 m_2 \over r}+\int_0^\infty  {dm \over k}  G_N{m \over k} 
{m_1 m_2e^{-mr} \over r}=G_N {m_1 m_2 \over r} \left(1+{1 \over k^2 r^2}
\right) \; .
\end{equation}
The KK contribution is suppressed at large distances over and above
that expected from having one additional
dimension, because of the amplitude suppression
near the brane. This is due to the barrier of the
analog quantum mechanical problem used to find the KK modes.

Now let us consider the opposite extreme. Suppose
we were at high energies and {\it suppose} it
were appropriate to use the plane wave form of the modes.
At a given location $y_0$, what would be
the corrections to Newton's law? They would
be
\begin{equation}
\label{newtonlarge}
V \sim G_N{m_1 m_2 \over r} + \int_0^\infty {dm \over k} 
G_N {m_1 m_2 e^{-mr}\over r} 
e^{3 k y_0} \sim G_N {m_1 m_2 \over r}  \left( 1+ {e^{3 k y_0} \over kr}
\right) \; .
\end{equation}

It is useful to write this answer in terms of mass
scales and compare to a flat five-dimensional space.
If $y_0$ is chosen to address the TeV hierarchy, one
finds the correction factor $(M_{Pl}/{\rm TeV})^3/kr$. Taking
$k\sim M_{Pl}$, one derives $M_{Pl}^2/{\rm TeV}^3 r$. With
a cross product background metric, one would
derive the TeV scale by choosing $r_c$ as $M^3 r_c =M_{Pl}^2$
where $M$ is of order a TeV and the mass of the KK modes
would be starting at $1/r_c$.
The corrections to gravity would be those of the number
of modes of energy less than $1/r$, which would  be $M_{Pl}^2/r M^3$.
This precisely agrees with the contribution one
would have in the warped background if one
saw the full continuum contribution.
This would of course be ruled out by current experiments
as it is far too strong a correction to gravity.

However, the calculation in the AdS
background based on the continuum form of the KK modes
is not appropriate. It helps to examine
the detailed form of the KK modes. 
Recall that they were derived in a background AdS
space in which there was a four-dimensional
flat brane with localized energy density.
One derives the KK modes by assuming they
factorize into momentum eigenstates with mass
$m$, where $m$ is determined by solving an analog
quantum mechanics problem describing the shape
of the KK mode in the fifth dimension. The
analog potential for these modes was dubbed
the ``volcano'' potential because there
was a delta-function at the origin, a barrier,
and then a smooth fall-off to zero
The zero mode is the single bound state.
All other modes are suppressed at the origin (as the
first calculation of corrections to Newton's law showed)
and then turn into continuum plane wave modes in the large
$y$ region, far from the brane.

\begin{figure}
\label{fig1}
\begin{center}
\epsfig{file=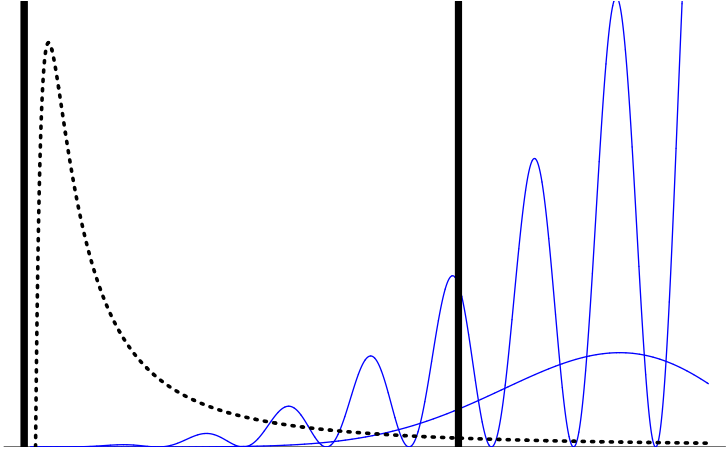,width=12cm}
\end{center}
\caption{Schematic view of the Kaluza-Klein gravity modes.
The x-axis is the fifth dimension. The left/right vertical
lines
represent the Planck/TeV branes. The ``volcano'' potential
rises then falls off rapidly away from the Planck brane.
Plotted are the squared amplitudes of two KK gravity modes
relative to the graviton zero mode. The heavy $m>>1$ TeV mode takes its
asymptotic (oscillating) form at the TeV brane, the other
mode exhibits the characteristic behavior for $m<<1$ TeV.
Very light modes with $m<10^{-4}$ eV would appear as flat lines,
since they track the zero mode.}
\end{figure}

At a given location $y_0$, modes which are sufficiently
light are suppressed relative to their continuum form,
while modes which have already assumed their asymptotic
form are unsuppressed. We can quantify this statement
by examining the explicit expression for the modes
Eq. (\ref{eigen}).
The asymptotic forms of the Bessel functions, and thus
the onset of continuum behavior, requires $mz_0$
much greater than 1, which is only true for modes
of mass greater than a TeV. This
is an important result. It says
that modes at all energies below
the strongly interacting regime 
are more suppressed than a continuum KK mode.
This result could have been anticipated
from Ref. \cite{rs2}, where
it was shown that quantization was in units
of approximately TeV. Modes
do not appear to have their continuum
form until they are at least this massive.

The suppression of the lighter modes is
addressed by looking at the asymptotic form
of the Bessel functions for small $mz$:
\begin{equation}
\label{smallz}
\sqrt{z} J_2(m z) \sim {m^2\over 8}z^{5/2}\; , 
~~~~ 
\sqrt{z} Y_2(m z) \sim -{4\over \pi m^2z^{3/2}} - {z^{1/2}\over\pi}\; .
\end{equation}
We see that $Y_2$ tracks the zero mode,
whereas $J_2$ rises sharply with respect to the
zero mode. So long as $Y_2$ dominates, the
contribution from the KK modes is as suppressed
relative to that of the zero mode as if
we were probing gravity on the Planck brane; e.g.
the corrections to Newton's law are given by Eqn. (\ref{newtonsmall}).
We find that $Y_2$ dominates so long
as we are exploring modes with mass less than
$1/(kz_0^2)$, which is approximately $10^{-4}$ eV.
All gravitational experiments to date
see the corrections to gravity to be as small
as if we were living on the Planck brane!

Modes with masses in the region intermediate
between $10^{-4}$ eV and 1 TeV are controlled by
the small $mz$ behavior of the dominant $J_2$ term.
If these modes had already reached their continuum
form at $z=z_0$, the cross section for
real emission of these modes would be proportional
to $E/({\rm TeV})^3$, where $E$ denotes the relevant
physical energy scale. This agrees with \cite{add2}
for $n=1$ extra dimensions, and leads to astrophysical
and collider effects which are clearly excluded by
observations. Using the actual form of these modes
at $z_0$, we find instead that the real emission
cross section is proportional to
\be
\label{sigemission}
\sigma \sim {E^6 \over({\rm TeV})^8} \quad .
\ee
So in fact the leading order energy dependence of these modes
agrees with the large torus compactifications of \cite{add1}
for the case of $n=6$ extra dimensions!
Because these effects are much softer in the infrared,
they turn out to be easily compatible with all
existing observations \cite{add2}.

In fact, a stronger result readily follows.
If matter is localized to any four-dimensional
flat brane between
the Planck and TeV branes, the force
between the matter will look four-dimensional
for energies less than a TeV. This means
that one could imagine doing physics in 
the bulk, analogous to what one
might have tried in the orbifold
case, to explain features of our observable world.

What emerges is a very compelling  picture. 
The world is five dimensional: the coordinate
$y$ extends to infinity. However,
for any observer localized to a given location $y_0$,
the modes of mass greater than $M_{Pl}\;$exp$-k|y_0|$ are
strongly coupled. The amplitude of lighter modes on the $y_0$ brane
is suppressed. Those of mass less than $1/kz_0^2$ ($10^{-4}$ eV
in our case) are coupled
by $1/M_{Pl}$ with further amplitude suppression.
Heavier modes are power law suppressed
over what would be expected had the metric been flat.
So the observer confined to the brane sees gravity
as essentially four-dimensional, no matter where
the brane is located! We can live
with an infinite extra dimension and simply
not know it.

The scenario presented here will be tested at future
collider experiments. To leading order in $(E/{\rm TeV})$,
real emission effects will mimic those of $n=6$ extra
dimensions in the scenario of \cite{add1}.
Because of the strong
power suppression, it is important to be able to probe
energy scales close to $(1/z_0)$. There may also be
detectable effects from virtual exchanges of KK modes.
However such effects are difficult to compute since
they are dominated by heavy modes near the TeV cutoff;
a string theory calculation is probably required to get
a reliable estimate.

This is a very tantalizing scenario.
It clearly ties in well with the holographic
picture advocated in \cite{verlinde}. Again,
with the infinite dimension, one expects
the gravitational theory to correspond
to a gauge theory cut-off in the ultraviolet.
Within this theory, there is a correspondence
between location $y$ and mass scale determined
by the shape of the zero mode. The
additional contribution of this paper is
to demonstrate that the Kaluza-Klein excitations
do not disturb this picture. They
give small corrections to the theory
of gravity, so long as one is at
sufficiently low energy.
This
new venue should provide new avenues
for addressing important problems
in cosmology and gravity.

\bigskip

{\bf Acknowledgements:} We wish to acknowledge
useful discussions with Savas Dimopoulos, 
Ann Nelson, Stuart Raby, Raman Sundrum, and
Herman Verlinde. We thank Martin Gremm
and Emanuel Katz for comments on the manuscript.
We also wish to thank
the Aspen Center for Physics, where this
work was initiated.
The research of Joe Lykken was supported by NSF
grant PHY94-07194, and by DOE grant DE-AC02-76CH03000. 
The research of Lisa Randall
was supported in part by DOE under cooperative
agreement DE-FC02-94ER40818 and
under grant number DE-FG02-91ER4071.

\end{document}